\documentclass[11pt,a4paper]{article}
\usepackage{graphicx}
\usepackage[T1]{fontenc}       
\usepackage[utf8]{inputenc}    

\usepackage{microtype}         

\usepackage[margin=2.5cm]{geometry}  

\usepackage{amsmath}
\usepackage{mathtools}         
\usepackage{amssymb}           
\usepackage{amsthm}            
\usepackage{bm}                
\usepackage{mathrsfs}          
\usepackage{dsfont}   
\usepackage[normalem]{ulem}
        
\usepackage{xcolor}            
\usepackage{float}             

\usepackage{tikz}
\usetikzlibrary{arrows,calc,positioning,external}
\usepackage[colorlinks, linkcolor=black, citecolor=black, urlcolor=black]{hyperref}
\usepackage[nameinlink,noabbrev]{cleveref}
\usepackage{caption}
\usepackage{subcaption}
\usepackage{wrapfig}
\usepackage{authblk}

\usepackage[backend=biber, style=alphabetic, maxbibnames=40, doi=false, isbn=false, url=false, eprint=false]{biblatex}
\theoremstyle{plain}
\newtheorem{theorem}{Theorem}

\newtheorem{corollary}{Corollary}[section]

\theoremstyle{remark}

\theoremstyle{definition}

\newtheorem{assumption}{Assumption}[section]

\newcommand{\numberset}{\mathbb}

\newcommand{\R}{\numberset{R}}

\DeclarePairedDelimiter{\norm}{\lVert}{\rVert}
\DeclarePairedDelimiter{\abs}{\lvert}{\rvert}

\DeclarePairedDelimiter{\tond}{(}{)} 
\DeclarePairedDelimiter{\quadr}{[}{]}
\DeclarePairedDelimiter{\graf}{\{}{\}} 
\DeclarePairedDelimiter{\scal}{\langle}{\rangle}

\DeclareMathOperator{\law}{law}

\newcommand{\eps}{\varepsilon}

\newcommand{\hess}{\nabla^2}

\newcommand{\expe}[1]{\mathbb{E}\quadr*{#1}}
\newcommand{\expewr}[2]{\mathbb{E}_{#1}\quadr*{#2}}

\newcommand{\pp}{\mathcal{P}}

\newcommand{\hatx}{\hat{X}}

\newcommand{\dd}{\, d}

\newcommand{\ff}{\mathcal{F}}

\title{Wasserstein mixing time of the unadjusted Langevin algorithm}
\author{Francesco Pedrotti}
\author{Peter A. Whalley}

\affil{ETH Z\"urich}

\begin{document}
\maketitle

\begin{abstract}
    We provide new estimates in Wasserstein distance for the asymptotic bias of the unadjusted Langevin algorithm, in the classical setting of log-smooth strongly log-concave measures. Our bound implies a Wasserstein mixing time of order $\kappa \sqrt{d}/\eps$, where $\kappa$ is the condition number, $d$ is the dimension, and $\eps$ is the target precision: this improves by a factor of $\sqrt{d}/\eps$ over the previous state-of-the-art results. 
\end{abstract}

\section{Introduction}

Sampling from a probability density $\pi \propto e^{-V}$ in $\R^d$ is a fundamental problem in a variety of fields.
	A popular family of methods to produce these samples, known as MCMC algorithms, consists in designing a stochastic process which is ergodic with respect to $\pi$, and simulating it for a long enough time.
	One  of the most fundamental choices is the \emph{Langevin dynamics (LD)}: given a (possibly) random initialization $\mu \in \pp_{2}\tond*{\R^d}$ and a standard $d$-dimensional Brownian motion $(B_t)_t$, it is  given by the solution of the SDE
	\begin{equation}
		\label{eq:LD}
		X_0\sim \mu, \quad dX_t = -\nabla V(X_t) dt + \sqrt{2}dB_t.
	\end{equation}
	It is well-known that, under mild assumptions on $V$, this stochastic process is ergodic with respect to $\pi$, making it a natural candidate for an MCMC algorithm (see \cite{PavBook}). Its practical implementation, however, requires considering a numerical scheme for \eqref{eq:LD}, such as the Euler--Maruyama method, which leads to the \emph{unadjusted Langevin algorithm (ULA)}, described by the easily implemented recursion 
	\begin{equation}
		\label{eq:ULA}
		\hatx_0 \sim \mu, \quad \hatx_{k+1} = \hatx_k - h\nabla V(\hatx_k) + \sqrt{2}\tond*{B_{(k+1)h} -B_{kh}},
	\end{equation}
	for a small step-size $h>0$. 
	
	Unfortunately, the above scheme is biased, which means that its limiting stationary distribution $\hat \pi$ is in general different from $\pi$; this leads to a trade-off in the choice of the step-size, with smaller values of $h$ leading to a smaller error in the bias, but to a higher required number of iterations of \eqref{eq:ULA} in order to converge to $\hat\pi$.
    \paragraph{Theoretical guarantees for ULA}
	In recent years,  considerable effort has been put into rigorously quantifying this trade-off \cite{Chewi26Book}, by proving error bounds in the approximation $\mu_k \coloneqq \law\tond*{\hatx_k} \approx \pi$ in different metrics, such as 
    the Wasserstein distance, defined for distributions $\mu,\nu \in \pp_2(\R^d)$ 
    \[
    W_2(\mu,\nu) = \inf_{(X,Y)\sim (\mu,\nu)} \sqrt{\expe{\abs*{X-Y}^2}},
    \]
    where the infimum runs over 
	all $\R^{d}$-valued random vectors 
	$X$ and $Y$
	defined on the same probability space with
	$\law(X) = \mu$ and $\law(Y) = \nu$.
    Our main result belongs to this line of work, and holds under the following classical assumption on the target measure.
    \begin{assumption}\label{ass:strong-log-conc-log-smooth}
		The potential $V\in C^2\tond*{\R^d}$ satisfies 
		\[
		\alpha I_d \preccurlyeq \hess V \preccurlyeq \beta I_d,
		\]
		for some $0<\alpha \leq\beta$. We denote the condition number by $\kappa \coloneqq \frac{\beta}{\alpha}$. 
	\end{assumption}

    \begin{theorem}\label{theorem:main}
    Under Assumption \ref{ass:strong-log-conc-log-smooth}, if  $h\le \frac1\beta$, we have
    \begin{equation}\label{eq:w2-bias}
        W_2\tond*{\pi, \hat\pi} \le  6 h \sqrt{\kappa \beta  d}.
    \end{equation}
    \end{theorem}
    Theorem \ref{theorem:main} immediately leads to convergence guarantees for the iteration complexity of ULA to achieve a desired accuracy in Wasserstein distance; more precisely, in this setting it is standard and more meaningful to work with the rescaled metric $\sqrt{\alpha}W_2$, in order to have scale-invariant results, cf. \cite[Remark 4.1.4]{Chewi26Book}.
    Under Assumption \ref{ass:strong-log-conc-log-smooth}, for $h\le \frac1\beta$ ULA converges exponentially fast to its biased limit, i.e. writing $\mu_k = \law(\hatx_k)$ we have $W_2(\mu_k,\hat \pi) \le (1-\alpha h)^{k}W_2(\mu,\hat{\pi})$.
    Thus, the iteration complexity to achieve a desired accuracy in $\sqrt{\alpha}W_2$-distance is essentially determined by the asymptotic bias and the step-size that it imposed, and 
    we deduce from Theorem \ref{theorem:main} the following Corollary.
    \begin{corollary}
        Under Assumption \ref{ass:strong-log-conc-log-smooth}, if  $h\le \frac1\beta$, we have
    \begin{equation}\label{eq:w2-convergence}
        W_2\tond*{\mu_k, \pi} \le \left(1-\alpha h\right)^{k} W_2(\mu,\hat{\pi}) + 6 h \sqrt{\kappa \beta  d}.
    \end{equation}
    In particular, for any $\eps>0$, choosing \[
    h = \frac{1}{\beta}\left(1 \wedge \frac{\eps}{12\sqrt{d}}\right)
    \]
    it suffices to take
    \begin{equation}\label{eq:complexity_bound}
    k \geq \kappa \tond*{1 \vee \frac{12\sqrt{d}}{\epsilon}}\log\left(1 + \frac{2\sqrt{\alpha}W_{2}(\mu,\hat{\pi})}{\epsilon}\right)
    \end{equation}
    iterations of ULA to ensure $\sqrt{\alpha}W_2(\mu_k,\pi) \le \eps$.
    \end{corollary}
   We remark that in \eqref{eq:w2-bias} the dependence on $h$ and $d$ can be shown to be optimal via explicit computations when the target is Gaussian; cf. \cite[Example 2]{wibisono2018sampling}. Additionally, the convergence time of \eqref{eq:ULA} towards its biased limit depends linearly on the condition number, already for anisotropic Gaussian targets. As a consequence, the linear dependence on the condition number in \eqref{eq:complexity_bound} is also optimal.

    \subsubsection*{Literature overview}
    Assumption \ref{ass:strong-log-conc-log-smooth} is classical in the study of the quantitative guarantees of the Unadjusted Langevin Algorithm (see \cite{dalalyan2017theoretical,durmus2017nonasymptotic} and the book \cite{Chewi26Book} for a more complete overview). Theoretical guarantees in Wasserstein distance under this assumption were first provided in \cite{durmus2019high,dalalyan2017further}, which establish the bound $W_2(\pi,\hat \pi) \lesssim \kappa\sqrt{hd}$. This bound was later improved to $W_2(\pi,\hat \pi) \lesssim \sqrt{\kappa hd}$ in \cite{durmus2019analysis}.
    Linear dependence in $h$ for the asymptotic bias was only established under additional assumptions on the third derivatives of $V$, but with a worse dependence on the dimension cf. \cite{durmus2019high,dalalyan2019user}. Subsequently, the dimension dependence was improved, but only under even stronger assumptions on the third derivatives of $V$ (see \cite{li2022sqrt,durmus2024asymptotic,dalalyan2026improved}).
    Besides Wasserstein distance, convergence guarantees have also been provided in other metrics, such as total variation, cf. \cite{dalalyan2017theoretical,durmus2017nonasymptotic}, and relative entropy, see \cite{vempala2023rapid,chewi2025analysis}, which also relaxes the log-concavity assumption to appropriate functional inequalities. Additionally, recent work \cite{Che-Che-Nil-Wea-2026} showed under suitable sparsity assumptions that dimension-free guarantees (up to logarithmic factors) can be obtained, but this crucially relies on replacing  the $2$-Wasserstein distance with an $\ell^\infty$-version.

    \section{Proof of Theorem \ref{theorem:main}}
    This section is devoted to the proof of Theorem \ref{theorem:main}. 
    We first prove the theorem under the additional regularity Assumption \ref{ass:extra-regularity} below, and remove it in Section \ref{sec:removing-regularity} with an approximation argument.
    \begin{assumption}\label{ass:extra-regularity}
        The potential $V$ is smooth and all of its partial derivatives of all orders have at most polynomial growth.
    \end{assumption}
    This additional assumption is considered for simplicity, to make rigorous and more transparent the application of standard semigroup and spectral theory in our argument. 
    For the convenience of the reader, we briefly recall some of these useful facts in the setting of Assumptions \ref{ass:strong-log-conc-log-smooth} and \ref{ass:extra-regularity}, and refer to the book \cite{bak-gen-led-2014}, the lecture notes \cite{kla-leh-2025} and the papers \cite{kla-put-2023-spec,kla-2023-log} for details.
    
    First, the Laplace operator $L$ associated to $\pi$
    is initially defined for $f \in C^\infty_c(\R^d)$ via
    \[
        Lf = \Delta f - \scal*{\nabla V, \nabla f}.
    \]
    From the definition and integration by parts, it follows that for $f,g \in C^\infty_c(\R^d) $
    \[
        \scal*{L f, g}_{L^2(\pi)} = - \int \scal*{\nabla f, \nabla g} \dd \pi,
    \]
    which shows that $-L$ is a symmetric and monotone operator. The operator $-L$
    is essentially self-adjoint, and in what follows we denote by $-L$ its minimal extension (with a mild abuse of notation), and by $D(L)$ its domain.
    Under our assumptions, $-L$ 
    has a discrete spectrum, 
    \[
        0 = \lambda_0 < \lambda_1 \le \ldots 
    \]
    and a corresponding orthonormal basis $(\varphi_k)_{k=0}^\infty$ in $L^2(\pi)$, where $\varphi_0 = 1$.
    
    The connection with the Langevin dynamics comes from the fact that $L$ can also be identified with the $L^2$-generator of the associated Markov semigroup $(P_t)_t$, which  is initially defined probabilistically for bounded measurable functions $f$ by
    \[
        P_t f(x) = \expe{f(X_t) \mid X_0 = x},
    \]
    and then extended to a contraction $P_t \colon L^2(\pi)\to L^2(\pi)$ by Jensen's inequality.
    Indeed, one can check that for $f \in D(L)$ we have 
    \[
        \frac{P_t f-f}{t} \to Lf \quad \text{ in } L^2(\pi) \text{ as } t\downarrow 0,
    \]
    and $D(L)$ is precisely the set of functions $f\in L^2(\pi)$ for which the left hand-side above admits a limit in $L^2(\pi)$.
    This suggests the identity $P_t f = e^{tL} f$, which holds in fact for all $f \in L^2(\pi)$.
    Here, $e^{tL} \colon L^2(\pi) \to L^2(\pi)$ is defined as an operator on $L^2(\pi)$, even if the operator $L$ is defined only on $D(L) \subset L^2(\pi)$: more generally, for every smooth $F\colon \R\to \R$ bounded on $(-\infty, 0]$, we can define
    \[
        F(L) \colon L^2(\pi) \to L^2(\pi)
    \]
    with spectral calculus, setting for $f\in L^2(\pi)$
    \[
        [F(L)]f  = \sum_{k =0}^\infty \scal*{f, \varphi_k}_{L^2(\pi)} F(-\lambda_k) \varphi_k.
    \]

    \subsection{Proof for regular distributions}\label{sec:extra-regularity}
    In this subsection, we prove Theorem \ref{theorem:main} under Assumption \ref{ass:extra-regularity}.
    Fix a step-size $0<h\le \frac1\beta$ and consider synchronous coupling of \eqref{eq:LD} and \eqref{eq:ULA} initialized at $X_0 = \hat X_0 \sim \pi$ (i.e. use the same driving Brownian motion).
    Setting $Z_k = X_{kh}-\hatx_k $, our aim is to prove that
    \begin{equation} \label{eq:desired-bound}
       \limsup_{k\to \infty} \sqrt{\expe{\abs*{Z_k}^2}} \le 6  h \sqrt{\beta \kappa d},
    \end{equation}
    from which the desired conclusion follows since $X_{t}\sim \pi$ for all $t\ge 0$ and $\hatx_k$ converges to $\hat \pi$ in $W_2$.
    Observe that after subtracting \eqref{eq:ULA} from \eqref{eq:LD}, $Z_k$ satisfies the recursion 
    \begin{equation}\label{eq:Zk}
        Z_0 = 0, \quad Z_{k+1} = (I_d-hA_k) Z_k + D_k, 
    \end{equation}
    where 
    \begin{equation}
        A_k \coloneqq \int_0^1 \hess V(\hatx_k+t Z_k) \dd t, \qquad  \qquad  \alpha I_d \preceq A_k \preceq \beta I_d,
    \end{equation}
    and 
    \begin{equation}     \label{eq:local_error}
        D_k\coloneq \int_0^h\{\nabla V(X_{kh})-\nabla V(X_{kh+r})\} \dd r,
    \end{equation}
    which we refer to as the local error at iteration $k$.
    \paragraph{Local error decomposition}
    The key idea is to decompose each local error \eqref{eq:local_error} into a telescoping increment and a martingale increment. At stationarity, sums of the local errors are expected to exhibit central-limit-theorem scaling \cite{gordin1969central,kipnis1986central,meyn2012markov}: their $L^2$-norm grows at rate $\sqrt{k}$, rather than at the naive rate $k$, where $k$ is the number of iterations (and summands). The decomposition makes this cancellation explicit, as the telescoping increments cancel across successive steps, whilst the conditionally centred martingale increments accumulate in $L^2$ at rate $\sqrt{k}$. We exploit both effects in the global error analysis below.
    
    We now derive the desired decomposition: denoting by $\mathcal{F}_t = \sigma\tond*{(X_s)_{0\le s \le t}}$ the $\sigma$-algebra generated by the diffusion up to time $t$, we claim that 
    there exist a vector field $g\in L^2(\pi;\R^d)$ and random vectors
    $M_{k+1}$ such that
    \begin{equation}
        D_k=g(X_{kh})-g(X_{(k+1)h})+M_{k+1},
        \qquad
        \mathbb{E}[M_{k+1}\mid\mathcal{F}_{kh}]=0,
        \label{eq:Dk-decomposition}
    \end{equation}
    and
    \begin{equation}
        \norm*{g}^2_{L^2(\pi)}\leq h^{2}\beta d,
        \qquad
        \mathbb{E}\abs{M_{k+1}}^2\leq 18\, h^3\beta^{2}d.
        \label{eq:Dk-decomposition-bounds}
    \end{equation}
    To achieve this, the idea is to take conditional expectations with respect to $F_{kh}$ in \eqref{eq:Dk-decomposition}, which gives
    \[
        (I-P_h)g(X_{kh}) = \expe{D_k\mid F_{kh}} = \int_0^h {(I-P_s) \nabla V (X_{kh})} \dd s.
    \]
    We can then identify $g$ by solving the discrete \emph{Poisson equation} \cite{meyn2012markov}
    \[
        (I-P_h) g = \int_0^h (I-P_s) \nabla V \dd s.
    \]
    More precisely, consider the function $G\in C^\infty(\R)$ defined by 
    \[
            G(x) = \begin{cases}
                \frac{h + (1-e^{hx})/x}{1-e^{hx}}\qquad &x \neq 0\\
                \frac{h}{2} \qquad &x = 0,
            \end{cases}
    \]
    which satisfies $\frac{h}{2}\le G(x) \le h$ for $x\le 0$.  
    We then define $g = G(L) \nabla V$ and observe that
    \begin{equation}\label{eq:L2-vector-field}
        \norm*{g}^2_{L^2(\pi)} \leq h^2\norm*{\nabla V}^2_{L^2(\pi)} = h^2\int \Delta V \dd\pi \leq h^{2}\beta d,
    \end{equation}
    which is the first bound in \eqref{eq:Dk-decomposition-bounds}. 
    Next, notice that 
    \[
    \begin{split}
        \expe{D_k \mid \mathcal{F}_{kh}} & = \int_0^h \tond*{\nabla V(X_{kh}) - \expe{\nabla V(X_{kh+s}) \mid \ff_{kh}} } \dd s  
        \\
        & = \int_0^h (I- P_s) \nabla V(X_{kh}) \dd s
        \\
        & = \int_0^h (I- e^{sL}) \nabla V(X_{kh}) \dd s
        \\
        & = \graf*{\quadr*{(I-e^{hL})G(L)}\nabla V}(X_{kh})
        \\
        & = g(X_{kh}) - P_h g(X_{kh})
        \\
        & = \expe{g(X_{kh}) - g(X_{(k+1)h})\mid \ff_{kh}}.
    \end{split}      
    \]
    Thus, defining $M_{k+1} \coloneq D_k - \tond*{g(X_{kh}) - g(X_{(k+1)h})}$ yields the desired decomposition in \eqref{eq:Dk-decomposition}.
    It remains to bound $\expe{\abs*{M_{k+1}}^2}$: to this end, we first bound
    
    \begin{align*}
        \expe{\abs*{g(X_{(k+1)h}) - g(X_{kh})}^{2}} &= 2\scal*{g,(I-e^{hL}) g}_{L^2(\pi)}
        \\
        & = 2\scal*{G(L)\nabla V,(I-e^{hL}) \tond*{G(L) \nabla V}}_{L^2(\pi)}
        \\
        & \le 2h^2 \scal*{\nabla V,(I-e^{hL}) { \nabla V}}_{L^2(\pi)}
        \\
        & = h^2\expe{\abs*{\nabla V(X_{(k+1)h}) -\nabla V(X_{kh}) }^2}
        \\
        & \leq 6h^3\beta^2 d.
    \end{align*}
    where we have used that 
    \begin{equation}\label{eq:displacement}
        \expe{\abs*{\nabla V(X_{kh+s}) -\nabla V(X_{kh}) }^2} \leq \beta^2 d\tond*{2s^2 \beta + 4s} \leq 6s\beta^2 d,
    \end{equation}
    for $0< s \leq \frac{1}{\beta}$. Additionally using \eqref{eq:displacement} we can also bound $\expe{\abs*{D_{k}}^2} \leq 3h^3\beta^2 d$, and hence using triangle inequality and \eqref{eq:Dk-decomposition} we have that $\expe{\abs*{M_{k+1}}^2} \leq 18 \, h^3 \beta^2 d$, which proves \eqref{eq:Dk-decomposition-bounds}.
    
    \paragraph{Global error bound}

    We can now insert decomposition \eqref{eq:Dk-decomposition} for $D_k$ in the recursion \eqref{eq:Zk} for $Z_k$: after rearranging, we deduce that
    \begin{equation}
        Z_{k+1}+g\tond*{X_{(k+1)h}} = (I_d -h A_k)\tond*{Z_{k}+g\tond*{X_{kh}}}+hA_k g(X_{kh}) +M_{k+1}
    \end{equation}
    Taking the square and expectation on both sides, we obtain 
    \begin{equation}
    \label{eq:shifted-recursion}
        \begin{split}
            \expe{\abs*{Z_{k+1}+g\tond*{X_{(k+1)h}}}^2} &\le (1-\alpha h)\expe{\abs*{Z_{k}+g\tond*{X_{kh}}}^2} + h\beta \,\expe{\abs*{g\tond*{X_{kh}}}^{2}} + \expe{\abs*{M_{k+1}}^{2}}\\
            &\le  (1-\alpha h)\expe{\abs*{Z_{k}+g\tond*{X_{kh}}}^2} + h^{3}\beta^{2}d + 18 \, h^{3}\beta^{2}d,
        \end{split}
    \end{equation}
    where we used \eqref{eq:Dk-decomposition-bounds}  and that
    \begin{equation*}
        \abs{(I-hA_{k})u+hA_{k}v}^2
        \leq(1-\alpha h)\abs{u}^2+h v^\top A_{k}v
    \end{equation*}
    for every $u,v\in\R^d$ since $h\beta\le 1$.
    From \eqref{eq:shifted-recursion} it follows that 
    \[
        \limsup_{k\to \infty} \expe{\abs*{Z_{k}+g\tond*{X_{kh}}}^2} \leq 19 \, h^{2} \kappa \beta d.
    \]
    Therefore, 
    \[
        \limsup_{k\to \infty} \sqrt{\expe{\abs*{Z_k}^2}} \le {\limsup_{k\to \infty} \tond*{\sqrt{\expe{\abs*{Z_{k}+g\tond*{X_{kh}}}^2}} +\sqrt{\expe{\abs*{g(X_{kh})}^2}}}} \leq 6 h \sqrt{\kappa \beta d},
    \]
    which is the desired bound \eqref{eq:desired-bound}.
    
    \subsection{Removing Assumption \ref{ass:extra-regularity}}
    \label{sec:removing-regularity}
    We finally remove the extra regularity assumption \ref{ass:extra-regularity}, thus concluding the proof of Theorem \ref{theorem:main}.
    For any small $\eps>0$, we denote by $\gamma_\eps \coloneqq \mathcal{N}(0, \eps I_d)$ the density of a Gaussian random variable with mean $0$ and covariance $\eps I_d$, and we set $\pi_\eps = \pi*\gamma_\eps$. Then, $V_\eps \coloneq - \log \pi_\eps \in C^\infty(\R^d)$ and 
    \[
        \frac{\alpha}{\alpha \eps + 1} I_d \preceq \hess V_\eps \preceq \beta I_d,
    \]
    see \cite{sau-wel-2014} and equation (6) in \cite{mik-she-2023} for the lower and upper bound respectively \cite{kla-put-2023-spec}.
    Moreover, Assumption \ref{ass:extra-regularity} is satisfied too, cf. \cite[Lemma 2.2]{kla-put-2023-spec}.
    For $h\le \frac1\beta$, we denote by $\hat{\pi}_{\eps}$ the biased limit of ULA with step-size $h$ and targeting $\pi_\eps$.
    By the triangle inequality, 
    \begin{align*}
        W_2(\pi,\hat \pi) & \le W_2(\pi,\pi_\eps) + W_2(\pi_\eps, \hat{\pi}_{\eps}) + W_2(\hat{\pi}_{\eps}, \hat \pi).
    \end{align*}
    An application of the previous step yields $W_2(\pi_\eps,\hat\pi_\eps) \le 6 h\sqrt{\tond*{1 + \alpha \eps}\kappa \beta d}$. Moreover, letting $Z\sim \gamma_\eps$, we have $W_2(\pi,\pi_\eps) \le \sqrt{\expe{\abs*{Z}^2}} \le \sqrt{\eps d}$.
    Thus, to conclude the proof, it suffices to show that $W_2(\hat \pi_\eps, \hat \pi) \to 0$ as $\eps \to 0$.
    To see this, fix any $0<\eps<\min{\tond*{1,\alpha^{-1}}}$, let $\hatx_0 =\hat Y_0 \sim \hat \pi$ and consider synchronous coupling for ULA targeting $\pi,  \pi_\eps$ respectively, i.e. 
    \begin{align*}
        \hat X_{k+1} &= \hat X_k - h \nabla V(\hat X_k) + \sqrt{2}\tond*{B_{(k+1)h} - B_{kh}}\\
        \hat Y_{k+1} &= \hat Y_k - h \nabla V_{\eps}(\hat Y_k) + \sqrt{2}\tond*{B_{(k+1)h} - B_{kh}},
    \end{align*}
    Defining $\tond*{\hat Z_{i}}_{i \in \mathbb{N}} = \tond*{\hat X_{i} - \hat Y_{i}}_{i\in \mathbb{N}}$, it follows by the bounds on $\hess V_{\epsilon}$ that
    \begin{align*}
        \sqrt{\expe{\abs{\hat Z_{k+1}}^{2}}} &\leq \tond*{1-\frac{\alpha h}{2}}  \sqrt{\expe{\abs{\hat Z_{k}}^{2}}} +  h\sqrt{\expewr{\hat \pi}{\abs{\nabla V_{\eps}  - \nabla V}^{2}}}.
    \end{align*}
    Taking the limit as $k \to \infty$ this implies
    \begin{align*}
        W_2(\hat \pi_\eps, \hat \pi) \leq \frac{2}{\alpha}\sqrt{\expewr{\hat \pi}{\abs{\nabla V_{\eps}  - \nabla V}^{2}}}.
    \end{align*}
    Observe now that $\nabla V_\eps \to \nabla V$ pointwise as $\eps\to 0$. Moreover, for $\eps$ small enough and some constant $C>0$ depending only on $\pi$, we have
    \[
        \abs*{\nabla V_\eps (x)-\nabla V(x)} \le C(1+\abs{x}) \in L^2(\hat \pi),
    \]
    since $\hat \pi$ is subgaussian \cite{vempala2023rapid}. Thus, by dominated convergence, we have 
    \[
        \sqrt{\expewr{\hat \pi}{\abs{\nabla V_{\eps}  - \nabla V}^{2}}} \to 0
    \]
    as $\eps \to 0$, which implies that $W_2(\hat \pi_\eps, \hat \pi) \to 0$ too, as desired.

    \subsubsection*{Acknowledgments}
    We would like to thank Yuansi Chen for inspiring discussions during preparation of another work, and acknowledge the use of ChatGPT 5.6 for complementary proof checking and literature exploration.
    
    \printbibliography 

    \appendix

\end{document}